\documentclass{jfm}

\usepackage{CJK}

\usepackage{ulem}
\usepackage[T1]{fontenc}
\usepackage{calligra}

\usepackage{graphicx}% Include figure files
\usepackage{dcolumn}% Align table columns on decimal point
\usepackage{bm}% bold math

\usepackage{times}
\usepackage{verbatim}
\usepackage{graphicx}
\usepackage{graphics,epsfig}

\usepackage{theorem}
\usepackage{makeidx}
\usepackage{amsmath}
\usepackage{epic}
\usepackage{amscd}

\usepackage{bbm}
\usepackage{xy}
\usepackage{amssymb}
\usepackage{epsfig}

\usepackage{cancel}
\usepackage{graphicx}
\usepackage{natbib}

\ifCUPmtlplainloaded \else
  \checkfont{eurm10}
  \iffontfound
    \IfFileExists{upmath.sty}
      {\typeout{^^JFound AMS Euler Roman fonts on the system,
                   using the 'upmath' package.^^J}%
       \usepackage{upmath}}
      {\typeout{^^JFound AMS Euler Roman fonts on the system, but you
                   dont seem to have the}%
       \typeout{'upmath' package installed. JFM.cls can take advantage
                 of these fonts,^^Jif you use 'upmath' package.^^J}%
      }
  \else
  \fi
\fi

\ifCUPmtlplainloaded \else
  \checkfont{msam10}
  \iffontfound
    \IfFileExists{amssymb.sty}
      {\typeout{^^JFound AMS Symbol fonts on the system, using the
                'amssymb' package.^^J}%
       \usepackage{amssymb}%
         
       \let\ge=\geqslant  
      }{}
  \fi
\fi

\ifCUPmtlplainloaded \else
  \IfFileExists{amsbsy.sty}
    {\typeout{^^JFound the 'amsbsy' package on the system, using it.^^J}%
     \usepackage{amsbsy}}
    {}
\fi

\newsavebox{\astrutbox}
\sbox{\astrutbox}{\rule[-5pt]{0pt}{20pt}}

\title[
Arbitrary-amplitude exact wave solutions: restricted superpositions
]{
On the exact solutions of (magneto)hydrodynamic systems and the superposition principles of arbitrary-amplitude helical waves
}

\author[J.-Z. Zhu
]%
{Jian-Zhou Zhu}
\affiliation{Su-Cheng Centre for Fundamental and Interdisciplinary Sciences, Gaochun, Nanjing 211316 Jiangsu, China and Li Xue Center, Gui Lin Tang Lab., 47 Bayi Cun, Yong'an, Fujian 366025 China
}

\pubyear{2010}
\volume{650}
\pagerange{119--126}
\date{?; revised ?; accepted ?. - To be entered by editorial office}
\begin{document}

\maketitle

\begin{abstract}
The principles of restricted superposition of circularly polarized arbitrary-amplitude waves for several hydrodynamic type models are illustrated systematically with helical representation in a unified sense. It is shown that the only general modes satisfying arbitrary-amplitude superposition to kill the generic nonlinearity are the mono-wavelength homochiral Beltrami mode and the one-dimensional-two-component stratified vorticity mode; while, there are other special superposition principles for some specific cases. We try to remark on the possible connections with the geo- and/or astro-physical fluid and magnetohydrodynamic turbulence issues, such as the rotating turbulence, dynamo and solar atmosphere turbulence, especially with the introduction of disorder locally frozen in some (randomly distributed) space-time regions. Recent disagreements about exact solutions of Hall and fully two-fluid magnetohydrodynamics are also settled down by such a treatment. This work complements, by studying the modes which completely kill the triadic interactions or the nonlinearities, previous studies on the thermalization purely from the triadic interactions, and in turn offers alternative perspectives of the nonlinearities.

\end{abstract}

\section{Introduction}
%\subsection{Restricted superposition principle of helical nonlinear waves in hydrodynamic type models}
Helical modes/waves are ubiquitous in a variety of natural systems:
\textit{\textbf{(I)}} For rotating fluids the inertial waves are ``transverse, circularly polarized and damped'' and ``are not limited to infinitesimal amplitudes'' (Chandrasekhar 1951, p. 86). The latter quote was iterated by Greenspan (1990, p. 187), saying that Chandrasekhar (1951) presented the helical waves as the ``exact solutions of the complete nonlinear equations,'' ``but the principle of superposition does not hold.'' Here, by ``the principle of superposition'' he meant \textit{arbitrary} superposition, which however raises questions such as what kind of special restricted superposition would be possible.
\textit{\textbf{(II)}} For the classical single-fluid magnetohydrodynamics (MHD) with an ambient field $\bm{B}_0$, it is well-known (Alfv\'en \& F\"althammer 1963) that the basic equations are solved by the so-called nonlinear Alfv\'en waves, as the linear dispersion relation requires $\bm{b}=\pm\bm{u}$ for the oscillating magnetic and velocity fields. We point out that it is this linear dispersion relationship that \textit{happens to} eliminate the nonlinear terms (and to allow arbitrary amplitude, thus the name ``nonlinear'', of the waves). So, such a nonlinear Alfv\'en wave is just an \textit{accident}; with other linear effects, such as the mean magnetic shear or rotation (see, e.g., Moffatt 1978), the dispersion relation no longer eliminates the nonlinearity, and the nonlinear wave solution then is not automatically given by the linear treatment, which highlights the importance of the dispersion relationship. This simple fact, that the traditional Alfv\'en wave does not solve rotating MHD (though $\bm{b}=\pm\bm{u}$ kills the nonlinearity), also indicates the arbitrary-amplitude nonlinear wave solutions are nontrivial, even though the arbitrary-amplitude flow (or ``vortex'', to distinguish from wave), such as the well-known Beltrami (1889) flow/vortex, is given; some of the seemingly familiar knowledge actually has deeper subtleties, especially after generalization and extension (to waves), and has some novel importance both in the formulation and results, especially for rotating MHD in comparison with the nonlinear classical Alfv\'en waves and for the extra insights about superposition of inertial waves (Chandrasekhar 1951; Greenspan 1990) crucial to rotating turbulence (see, e.g., Cambon \& Jacquin 1989; Waleffe 1993; Chen et al. 2005). These vortex and waves are fundamental to rotating (magneto)hydrodynamic turbulence (see, e.g., Sagaut \& Cambon 2008; Davidson 2013).
\textit{\textbf{(III)}} For the principle of superposition of helical (cyclotron) waves in more complicated magnetofluid models with richer physics, such as the Hall and fully two-fluid MHD, to form the exact solutions, recently there appeared some controversies: Mahajan \& Krishan (2005) initiated a paper on the exact wave solutions for Hall MHD, and, Mahajan \& Miura (2009) proposed linear superposition of nonlinear waves in (fully) two-fluid MHD. Both these were partially disagreed by Galtier (2006) and Verheest (2009), which of course also challenges the relevance made for astrophysical (such as the solar atmosphere) turbulence (but we will show that astrophysical turbulence relevance made earlier might still be salvaged and updated by considering the possible localized disorder in the system). Such superposition issue wants further elucidation for pure theoretical understanding of the dynamics and for recognizing flow and field structures in nature. To clarify the arbitrary-amplitude wave superposition issue for all these situations, %among others purposes of astrophysical illumination,
a more systematic treatment with the appropriate tool is thus called for.
%At a fundamental but transparent level,
\textit{In a previous study (Zhu, Yang \& Zhu 2014), we offered the results of the thermalization effects purely from the triadic interactions of the helical modes representing the nonlinear interaction properties (detailed conservation laws, Liouville theorem et al.) for a variety of hydrodynamic type systems, and here we will complement it by studying the superposed arbitrary-amplitude helical waves which completely kills the triadic interactions or the nonlinearities, to offer alternative perspectives of the nonlinearities in turn.}

The core of the discussion of superposition of helical waves, each of whom presents alignment of its curl and itself, is whether and how the nonlinearities would be eliminated after superposition.
%Such thoughts of special exact solutions with alignment of ``flux lines'' and ``vortex lines'' (Beltrami, 1889) is not new.
An example is the Beltrami (1889) flow and force-free equilibria in astrophysics and laboratory fusion communities. [To put in other words, ABC flow (Beltrami 1889; Dombre et al. 1986), as a particular example of Beltrami field, is the superposition of three helical modes of the same wavelength and of the same polarity, or the superposition of three ``chiroids'' of the same chirality as termed by Zhu, Yang \& Zhu (2014) for better intuition.]
%However, as said in the first paragraph, though certain flow kills the nonlinearity it does not simply ensure a wave solution.
Yet another probably less popular example is the ``stratified vorticity mode'' termed by Moses (1971). We thus will revisit and generalize Beltrami and Moses' results, and, show that \textit{there are actually no other generic exact solutions}, with the help of helical representation which have proved to be powerful (Moses 1971, Cambon \& Jacquin 1989; Waleffe 1992; Waleffe 1993), which in turn enables us to effectively revisit the arbitrary-amplitude superposition of fundamental inertial waves, Alfv\'en waves and plasma cyclotron waves, and to obtain some further illumination about various flows.

%Here we\\
%\subsection{Helical representation in Fourier space}
The above discussions can be made more definite, and the point more concrete, as follows.
%\subsection{Helical mode representation and special superpositions}
Moses (1971) has used the complete and orthogonal eigenfunctions of the curl operator for expansion of a vector function, a field not necessarily incompressible. And ``the original Helmholtz theorem has been sharpened in two ways. First, two irrotational vectors have been introduced in the decomposition of a general vector, each of which is the curl of its own vector potential. Second, a procedure has been given for obtaining the vector and scalar potentials''. Here the ``Helmholtz theorem'' is what decomposes a field into the rotational/transverse and irrotation/longitudinal parts, representable respectively by the curl of a potential vector and the gradient of a scalar potential. The longitudinal component of the field is represented with the eigenfunctions corresponding to the eigenvalue $0$ and is uncoupled from the transverse part.

Then, for a 3D transverse vector field $\bm{v}(\bm{r})$ the helical mode/wave representation in Fourier space reads (Moses 1971)
\begin{equation}\label{eq:FourierHelical}
\bm{v}(\bm{r})%=\sum_{\bm{k}} \hat{\bm{v}}(\bm{k})e^{\hat{i}\bm{k}\cdot \bm{r}}
=\sum_c \bm{v}^c(\bm{r})=\sum_{\bm{k},c} \hat{\bm{v}}^c(\bm{k}) e^{\hat{i}\bm{k}\cdot \bm{r}}=\sum_{\bm{k},c} \hat{v}^c(\bm{k})\hat{\bm{h}}_c(\bm{k})e^{\hat{i}\bm{k}\cdot \bm{r}}.
\end{equation}
Here $\hat{i}^2=-1$ and $c^2=1$ for the chirality indexes $c=$ ``+'' or ``-''. The helical mode bases (complex eigenvectors of the curl operator) have the following properties $\hat{i}\bm{k}\times \hat{\bm{h}}_c(\bm{k})=ck\hat{\bm{h}}_c(\bm{k})$, $\hat{\bm{h}}_c(\bm{-k})=\hat{\bm{h}}_c^*(\bm{k})=\hat{\bm{h}}_{-c}(\bm{k})$ and $\hat{\bm{h}}_{c_1}(\bm{k})\cdot\hat{\bm{h}}_{c_2}^*(\bm{k})=2\delta_{c_1,c_2}$ (Euclidean norm). $\hat{\bm{h}}_c(\bm{k})e^{\hat{i}\bm{k}\cdot \bm{r}}$ is the eigenfunction of the curl operator corresponding to the eigenvalue $ck$. Or, with the case $c=0$ also included for the compressible field, the variable $c$ ``itself may be considered to be the eigenvalue of the operator $(-\nabla^2)^{-1/2}\nabla\times$ when this operator is properly interpreted.'' The bases can be simply constructed as (Greenspan 1990; Waleffe 1992)
\begin{equation}\label{eq:h}
\hat{\bm{h}}_c(\bm{k})=(c\hat{i}\bm{l}+\bm{l}\times\bm{k}/k)/l,
\end{equation}
with $\bm{l}$ being perpendicular to $\bm{k}$. The structure $\hat{\bm{h}}_c(\bm{k})e^{\hat{i}\bm{k}\cdot\bm{r}}$ is common in inertial waves of rotating fluids and cyclotron waves of plasmas, being circularly
polarized, with $c=\pm$ representing opposite chirality.

Then, for a three-dimensional (3D) transverse vector field $\bm{v}$ the helical mode/wave representation in Fourier space reads (Moses 1971; Waleffe 1992)%\citep{Moses71,w92}
\begin{equation}\label{eq:FourierHelical}
\bm{v}%=\sum_{\bm{k}} \hat{\bm{v}}(\bm{k})e^{\hat{i}\bm{k}\cdot \bm{r}}
=\sum_c \bm{v}^c=\sum_{\bm{k},c} \hat{\bm{v}}^c(\bm{k}) e^{\hat{i}\bm{k}\cdot \bm{r}}=\sum_{\bm{k},c} \hat{v}^c(\bm{k})\hat{\bm{h}}_c(\bm{k})e^{\hat{i}\bm{k}\cdot \bm{r}}.
\end{equation}
Here $\hat{i}^2=-1$ and $c^2=1$ for the chirality indexes $c=$ ``+'' or ``-''. [For consistency of notation, every complex variable wears a hat and its complex conjugate is indexed by ``*''.] For convenience, we normalize the box to be of $2\pi$ period and that $k\ge 1$. The helical mode bases (complex eigenvectors of the curl operator) have the following properties $$\hat{i}\bm{k}\times \hat{\bm{h}}_c(\bm{k})=ck\hat{\bm{h}}_c(\bm{k}),$$ $$\hat{\bm{h}}_c(\bm{-k})=\hat{\bm{h}}_c^*(\bm{k})=\hat{\bm{h}}_{-c}(\bm{k})$$ and $\hat{\bm{h}}_{c_1}(\bm{k})\cdot\hat{\bm{h}}_{c_2}^*(\bm{k})=\delta_{c_1,c_2}$ (Euclidean norm). $\hat{\bm{h}}_c(\bm{k})e^{\hat{i}\bm{k}\cdot \bm{r}}$ is the eigenfunction of the curl operator corresponding to the eigenvalue $ck$. Or, with the case $c=0$ also included for the compressible field, the variable $c$ ``itself may be considered to be the eigenvalue of the operator $(-\nabla^2)^{-1/2}\nabla\times$ when this operator is properly interpreted.'' (Moses 1971) %\citep{Moses71}
The bases can be simply constructed as (Chandrasekhar 1951; Greenspan 1990; Waleffe 1992) %\citet{Greenspan90,w92}
$$\hat{\bm{h}}_c(\bm{k})=(c\hat{i}\bm{p}+\bm{p}\times\bm{k}/k)/(\sqrt{2}p),$$ with $\bm{p}$ being perpendicular to $\bm{k}$. The structure $\hat{\bm{h}}_c(\bm{k})e^{\hat{i}\bm{k}\cdot\bm{r}}$ is common in inertial waves of rotating fluids and cyclotron waves of plasmas, being circularly
polarized, with $c=\pm$ representing opposite chirality. Such a tool has recently been used in direct numerical simulations of homochiral Navier-Stokes \citep{bmt12}. For better or alternative physical intuition, we may conveniently call (Zhu, Yang \& Zhu 2014) $$\breve{\bm{v}}^c(\bm{r}|\bm{k})=\hat{v}^c(\bm{k})\hat{\bm{h}}_c(\bm{k})e^{\hat{i}\bm{k}\cdot \bm{r}}+c.c.,$$ with $c.c.$ for ``complex conjugate'', a ``chiroid'' which is maximally/purely helical or of highest degree of chriality, since the helicity contribution of it is (Zhu 2014) $$\nabla\times\breve{\bm{v}}^c(\bm{r}|\bm{k})\cdot\breve{\bm{v}}^c(\bm{r}|\bm{k})=2ck|\hat{v}^c(\bm{k})|^2=ck|\breve{\bm{v}}^c(\bm{r}|\bm{k})|^2.$$
%other corresponding chemistry terminologies, such as enantiomer, enantiopure and racemic etc. may also be tentatively borrowed for this purpose.

As the above discussions indicate, it is very much with an eye to understand turbulence that this investigation about exact solutions and superposition principles is pursued. In particular, we have it in mind that the exact solution(s) found here could be the stable mode(s), working as the persistent structure(s) of the turbulence, or, even though unstable, statistical mechanics may favor such state(s), just as the Beltrami field/flow (Moffatt 1986; Galloway \& Frisch 1987) which may be favored statistically as a one-chiral-sector-dominated state (OCSDS) of condensation (Zhu, Yang \& Zhu 2014). And, finally, we need to point out in the end of this introductory discussion that this note is cross-disciplinary, covering a wide range of models, with a unified treatment; so, it is necessary to summarize the main results to be communicated below for an easy reading: We first show for the first time the ``uniqueness" of two exact arbitrary-amplitude/nonlinear modes, then for each model we discuss explicit examples for them and, when possible, new specific exact nonlinear modes. Remarks on possible relevance to turbulence, with new ideas, and some clarifications of existing work will also be offered.

\section{Generic exact nonlinear modes}
\subsection{Two exact modes for the Navier-Stokes equations}
Now we consider how the superposition of the ``arbitrary-amplitude'' helical modes can eliminate $$\nabla\times(\bm{u}\times\bm{v}),$$ with $\bm{u}$ being the velocity field, which is the nonlinearity in the typical dynamics studied here. We are looking for the situation where $\bm{u}\times\bm{v}$ is the gradient of a \textit{potential function} so that the nonlinearity vanishes.
%One interesting result from the above is the general representation of the Beltrami field. To see this, we carry out the direct algebra
For Navier-Stokes, $\bm{v}$ is the vorticity $\bm{\Omega}=\nabla\times\bm{u}$, and from $$\nabla\times\bm{u}=\sum_{\bm{k},c} \hat{u}^c(\bm{k})[\hat{i}\bm{k}\times\hat{\bm{h}}_c(\bm{k})]e^{\hat{i}\bm{k}\cdot \bm{r}}=\sum_{\bm{k},c} ck\hat{u}^c(\bm{k})\hat{\bm{h}}_c(\bm{k})e^{\hat{i}\bm{k}\cdot \bm{r}},$$
we have:
\begin{enumerate}
  \item mono-wavelength homochiral Beltrami mode: when $ck$ is constant% or $\hat{\bm{h}}_c(\bm{k})$ is a constant vector
  , $\nabla\times\bm{u}=ck\bm{u}$ and the corresponding potential function is some constant.
A most celebrated example is the ABC flow.%:
%  \begin{eqnarray}\label{eq:ABC}
%% \nonumber to remove numbering (before each equation)
% \nonumber \dot{x} &=&u_x= A\sin(z)+B\cos(y) \\
%  \dot{y} &=&u_y= C\sin(x)+A\cos(z),\\
% \nonumber \dot{z} &=&u_z= B\sin(y)+C\cos(x)
%\end{eqnarray}
%with $A$, $B$ and $C$ being the real numbers.
  \item ``stratified vorticity mode'' of Moses (1971): when all $\bm{k}$s are parallel, with $\bm{k}/k=\pm\bm{\kappa}$ and $\bm{\kappa}$ being a fixed but arbitrary vector, $\nabla\times\bm{u}$, as $\bm{u}$, is perpendicular to $\bm{\kappa}$; so, $\bm{u}\times(\nabla\times\bm{u})$ is a vector function in the direction along $\bm{\kappa}$, depending only on the variable $\bm{r}\cdot\bm{\kappa}$ with the curl being zero (or, alternative, the corresponding potential function being the antiderivative of this function.)
%To be definite, here we also give a simple periodic example flow composed of three, or even more, modes with $\bm{\kappa}=(0,0,1)$:
%\begin{eqnarray}\label{eq:XYz}
%% \nonumber to remove numbering (before each equation)
% \nonumber \dot{x} &=&u_x= X_1\cos(k_1z)+X_2\cos(k_2z)+X_3\cos(k_3z)+... \\
%  \dot{y} &=&u_y= Y_1\cos(k_1z)+Y_2\cos(k_2z)+Y_3\cos(k_3z)+...,\\
% \nonumber \dot{z} &=&u_z= 0
%\end{eqnarray}
%with $X_i$, $Y_i$ and $k_i$ ($i=1,2,3,...$) being real numbers. [Of course, $u_x$ and $u_y$ in general can be arbitrary functions of $z$ and may have Kelvin-Helmholtz type of instability.]
And, without loss of generality one can take $\bm{\kappa}=(0,0,1)$ such that the flow has only the $x$ and $y$ components depending merely on the $z$ coordinate, and one may conveniently call such a one-dimensional-two-component (1D2C) mode the XYz flow (see later discussions.)
\end{enumerate}

Since we are considering the ``arbitrary-amplitude'' modes, which excludes some specific flows such as, with $\alpha(\bm{r})$ being $\bm{r}$-dependent instead of a constant, $\nabla\times\bm{u}=\alpha(\bm{r})\bm{u}$ which imposes tenuous relationships between the amplitudes of the Fourier modes, it appears to be no other generic superpositions to make $\bm{u}\times(\nabla\times\bm{u})$ an exact differential, which will be shown in the next subsection. Time dependency, damping or wavy, can be accounted by multiplying the corresponding factor $e^{-\hat{i}\omega t}$ which is determined with vanishing nonlinearity, formally the same as the derivation of infinitesimal amplitude linear waves. \textit{Thus the above \textit{(a)} and \textit{(b)} are the only generic superposition principles of nonlinear wave solutions.
For some special combinational structures of the nonlinear terms in other more complex models, there can be other types of ``accidental'' solutions due to the mutual cancelations between the nonlinear terms}, such as the Alfv\'en wave (without any restriction on the superposition) in single-fluid MHD with an ambient field, as discussed earlier (there exist also such nonlinear Alfv\'en waves in the $\bm{k}$ plane perpendicular to the rotating axis even for rotating MHD, constituting \textit{a third superposition principle}.) Note in particular that the classical nonlinear Alfv\'en wave can be superposed over arbitrary $\bm{k}$ and thus can be of ``arbitrary'' shape (to be mathematically rigor, in $L^2$ space.)

\subsection{`Uniqueness' of the two exact modes for the generic nonlinearity}
We first show here that \textit{(a)} mono-wavelength homochiral Beltrami mode and \textit{(b)} stratified vorticity mode as discussed in the last section are the only generic exact solutions of the basic Navier-Stokes equations. This can be accomplished by letting all other helical modes' amplitudes be zero (as they are ``arbitrary''), except for two that do not belong to the same group of the above two categories; these two modes, with arbitrary amplitudes, do not kill the nonlinearity. The simplest way probably is to work in the Fourier space where the equation is given in the symmetric form in $\bm{p}$ and $\bm{q}$ (Waleffe 1992) as follows
\begin{equation}\label{eq:NS}
    2\partial_t \hat{u}_{c_{\bm{k}}}=
    \frac{1}{2}\sum_{\bm{k}=\bm{p}+\bm{q}}\sum_{c_{\bm{p}},c_{\bm{q}}}  (c_{\bm{q}}q-c_{\bm{p}}p)\hat{\bm{h}}_{c_{\bm{p}}} \times \hat{\bm{h}}_{c_{\bm{q}}} \cdot \hat{\bm{h}}^*_{c_{\bm{k}}} \hat{u}_{c_{\bm{p}}}\hat{u}_{c_{\bm{q}}}.
\end{equation}
[For notational convenience, here and sometimes later, we denote the chirality coming with $\bm{k}$ with $c_{\bm{k}}$ and thus $\hat{\bm{h}}_{c_{\bm{k}}}$, and similarly for those with $\bm{p}$ and $\bm{q}$.] So, when there are only two modes of wavevector $\bm{p}$ and $\bm{q}$, the right hand side becomes $(c_{\bm{p}}p-c_{\bm{q}}q)\hat{\bm{h}}_{c_{\bm{p}}} \times \hat{\bm{h}}_{c_{\bm{q}}} \cdot \hat{\bm{h}}^*_{c_{\bm{k}}} \hat{u}_{c_{\bm{p}}}\hat{u}_{c_{\bm{q}}}$ with $\bm{k}=\bm{p}+\bm{q}$. Since $\hat{u}_{c_{\bm{p}}}\hat{u}_{c_{\bm{q}}}$ can be an arbitrary number, the nonlinear term can generically vanish only when the coefficient before $\hat{u}_{c_{\bm{p}}}\hat{u}_{c_{\bm{q}}}$ is zero. That is,
\begin{itemize}
  \item `\textit{(a)}', $\bm{p}$ and $\bm{q}$ are of the same wavelength and chirality with $c_{\bm{q}}q-c_{\bm{p}}p=0$;
  \item and, `\textit{(b)}', $\hat{\bm{h}}_{c_{\bm{p}}} \times \hat{\bm{h}}_{c_{\bm{q}}} \cdot \hat{\bm{h}}^*_{c_{\bm{k}}}=0$.
\end{itemize}

For `\textit{(b)}', note that $\bm{-k}$, $\bm{p}$ and $\bm{q}$ are in the same plane, forming a triangle, so in Eq. (\ref{eq:h}) we can let $\bm{l}=\bm{p}\times\bm{q}$ for nonvanishing $\bm{p}\times\bm{q}$, \textit{i.e.}
\begin{equation}\label{eq:hk}
\hat{\bm{h}}_{c_{\bm{k}}}=\frac{c_{\bm{k}}\hat{i}\bm{p}\times\bm{q}}{|\bm{p}\times\bm{q}|}+\frac{\bm{p}\times\bm{q}\times\bm{k}}{|\bm{p}\times\bm{q}|k},
\end{equation}
and, similarly $\hat{\bm{h}}_{c_{\bm{p}}}$ and $\hat{\bm{h}}_{c_{\bm{q}}}$, which when brought into $\hat{\bm{h}}_{c_{\bm{p}}} \times \hat{\bm{h}}_{c_{\bm{q}}} \cdot \hat{\bm{h}}^*_{c_{\bm{k}}}=0$ to write down the explicit expressions of the real and imaginary parts, with every $\bm{k}$ replaced by $\bm{p}+\bm{q}$, immediately lead to $(c_{\bm{k}}+c_{\bm{p}}+c_{\bm{q}})\bm{p}\times\bm{q}=0$, thus $\bm{p}\times\bm{q}=0$ (contrary to the starting assumption), i.e., $\bm{p}$ and $\bm{q}$ must be on the same direction. Since $\bm{p}$ and $\bm{q}$ are of any two modes, we have proved that the nonlinearity generically vanishes for and only for the mono-wavelength homochiral Beltrami mode or the stratified vorticity mode.
Note also that obviously time dependency, damping (say, by viscosity) or wavy (due to, say, rotation), can be accounted by multiplying the corresponding factor $e^{-\hat{i}\omega t}$. We determine $\omega$ with vanishing nonlinearity, formally the same as the derivation of infinitesimal amplitude linear waves. Thus we have shown that the above \textit{(a)} and \textit{(b)} are the only generic superposition principles of nonlinear wave solutions.

The fact that either of the above two modes leads to vanishing nonlinearity (Beltrami 1889; Moses 1971; Moffatt 1978; Dombre et al. 1986; Mahajan \& Krishan 2005; Verheest 2009) and no coupling in the kinetic energy and helicity equations (see, e.g., Galtier 2003, for weak turbulence theory; and references therein) has been well-known, but we have shown that it is necessary to have either or both of them for eliminating the generic nonlinearity; that is, \textit{there are no others and they are the ``unique'' generic modes}. Note that the above `proof' for the generic superposition principle is only for the simplest Navier-Stokes equations, which does not exclude other types of \textit{``accidental'' solutions} due to the mutual cancelations between the nonlinear terms in more complex systems, such as the Alfv\'en wave (without any restriction on the superposition) in single-fluid MHD with an ambient field as discussed earlier.

We will apply the above fundamental considerations, with necessary extensions, to construct the exact nonlinear wave solutions of rotating (magneto)fluids, and clarifications of the controversial issues about Hall and two-fluid MHD exact solutions will also be offered in this light as the byproducts. Other special superposition principles will also be discovered. The purpose is to clarify the nontrivial superposition principle of nonlinear problems and to find insights into nonlinear dynamics relevant to natural situations (dynamo, turbulence etc.): As said, the exact solution(s) found here, just as the Beltrami field/flow (Moffatt 1986; Galloway \& Frisch 1987), could be the stable mode(s), or, even though unstable, may be favored statistically by an isotropic 3D OCSDS of condensation for natural chirality amplification or imposed/restricted chirality cancelation (Zhu, Yang \& Zhu 2014; Zhu 2014). Note that since we are treating the issues in a unified way, some of the explicit formulae, especially the examples, when carried from one case to another with possible trivial extensions, won't be repeated except for some brief remarks.

\section{Exact wave solutions of various hydrodynamic type systems: Rotating neutral fluids, rotating classical single-fluid MHD and two-fluid MHD}

\subsection{Rotating neutral fluids}\label{app:ABC}
The vorticity $\bm{\Omega}=\nabla\times\bm{u}$ equation of an ideal fluid in a frame rotating at $\bm{\Omega}_0$ reads
\begin{equation}\label{eq:rf}
    \partial_t %\nabla\times\bm{u}
    \bm{\Omega}=\nabla\times(\bm{u}\times%(\nabla\times\bm{u})
    \bm{\Omega})+2(\bm{\Omega}_0\cdot\nabla)\bm{u}%-\nu\Delta\bm{u}
    .
\end{equation}
The two types of exact solutions are the following:

\subsubsection{The mono-wavelength homochiral Beltrami mode}
The mono-wavelength homochiral mode is defined by
\begin{equation}\label{eq:mhU}
\bm{u}=\sum_{|\bm{k}|=k_*,c=c^*} \hat{u}^c(\bm{k})\hat{\bm{h}}_c(\bm{k})e^{\hat{i}\bm{k}\cdot \bm{r}} e^{-\hat{i}\omega_c(\bm{k}) t} %e^{-\nu k_*^2t}%,\ \bm{b}=\gamma\bm{u},
\end{equation}
where $\omega_c=c2\bm{k}\cdot\bm{\Omega}_0/k$ is determined by the linear parts of Eq. (\ref{eq:rf}), same as that of the conventional rotating-fluid dispersion relation of infinitesimal wave perturbations; $k_*$ is a fixed but positive arbitrary number; $c^*$ is fixed but can be either $+$ or $-$. A damping factor $e^{-\nu k_*^2t}$ can be trivially included, without changing the superposition property, when viscosity is considered (Greenspan 1990). We iterate that Chandrasekhar (1951) noted that the single helical wave kills the nonlinearity and thus can be of arbitrary amplitude, but in general the superposition, other than the principles we are clarifying with helical representation, of waves only hold for infinitesimal-amplitude linear ones.

This ``restricted'' superposition is in fact a quite general superposition of inertial waves, the only constraints being that $c^*$ is constant and $\bm{k}$ lies on the sphere with radius $k_*$. We may even have an infinite number of modes and a continuous distribution of $\hat{u}(\bm{k})$ on this sphere. One may find it interesting to see that, in the large-$k$ or continuous-$\bm{k}$ limit, this exact solution has a very rich temporal structure with arbitrary spectrum of $\omega_c(\bm{k})\in [-2, 2]\Omega_0$. Such an exact (damping) wave solution comes from a superposition of inertial waves of arbitrary amplitudes. Thus, they have different propagating speeds (fastest in the direction of rotation) but same damping rate. We may call them the Beltrami waves; indeed, when the rotating speed and viscosity vanish, the solution is the generalization of the well-known ABC flow (Beltrami 1889, Dombre 1986), which can be made more explicit as follows.

%\subsubsection{The generalized ABC flows}\label{app:ABC}
For example, consider a $2\pi$-periodic field with $k=5$, we have modes with $k_l=\pm 3, \ k_m=\pm 4, \ k_n=0$, besides those ABC modes with $k_l=\pm 5, \ k_m=0, \ k_n=0$, with $l$, $m$ and $n$ permutating among $x$, $y$ and $z$. We can easily write down the equations for such a flow. For example, we can add one more helical component from the contribution of the two modes $(3, 4, 0)$ and $(-3,-4,0)$, $[-4\sin(3x+4y),3\sin(3x+4y),5\cos(3x+4y)]$ %constructed by the inverse of helical wave decomposition (or by appropriate transformation/rotation from any of the ABC ones),
added to the ABC flow and have the following ABC$d$ flow
%(for brevity, the factor from $e^{-\hat{i}\omega(\bm{k}) t}$ of each component is omitted):
\begin{eqnarray}\label{eq:ABCD}
 \nonumber \dot{x} &=&u_x= A\sin(5z)+B\cos(5y)-4d\sin(3x+4y) \\
  \dot{y} &=&u_y= C\sin(5x)+A\cos(5z)+3d\sin(3x+4y)\\
 \nonumber \dot{z} &=&u_z= B\sin(5y)+C\cos(5x)+5d\cos(3x+4y)
\end{eqnarray}
which can be made to be more symmetric with further permutations, and thus the ABC$def$ flow,
%\lipsum[1]
%\begin{widetext}
%\begin{minipage}
% \begin{onecolumn}
\begin{eqnarray}\label{eq:ABCDEF}
 \nonumber \dot{x} =u_x= A\sin(5z)+B\cos(5y)%\nonumber\\
 -4d\sin(3x+4y)+5e\cos(3y+4z)+3f\sin(3z+4x) \\
 \nonumber \dot{y} =u_y= C\sin(5x)+A\cos(5z)%\nonumber\\
  +3d\sin(3x+4y)-4e\sin(3y+4z)+5f\cos(3z+4x). \\
 \nonumber \dot{z} =u_z= B\sin(5y)+C\cos(5x)%\nonumber\\
 +5d\cos(3x+4y)+3e\sin(3y+4z)-4f\sin(3z+4x)
\end{eqnarray}
% \end{onecolumn}
%\end{minipage}
%\end{widetext}
%\lipsum[1]
And, even more modes from (4,3,0), (3,-4,0), (-4,3,0) and their permutations can be added. In total, we can have $15$ components constructed from the modes on this shell. There are concerns that the traditional ABC flow might not have enough chaotic region (see, e.g., Galloway 2012) for a fast dynamo, thus the above generalization with more modes would intuitively be expected to be potential to be more chaotic and to improve: It is just the general mathematical knowledge that in general systems in higher dimension of freedoms tend to be more chaotic. Actually, when one of the parameters of ABC flow is set to be zero, the flow becomes integrable and that no chaos, which was shown by Dombre et al. (1986), whose Painleve analysis ``suggests that the ABC flows present no further cases of integrability, other than the obvious ones [with A or B or C being set to zero]'', the extrapolation of which is our intuitive expectation. At least the ABC$d$ flow appears to be a promising start for numerical clarification of this point.

Coming back to Beltrami waves with $\omega=\omega_c(\bm{k})$, we see that when $A=e=f=0$ in the above and when $\bm{\Omega}_0$ is identified to be parallel to $\bm{z}$, the flow is independent of $z$ and then becomes the so-called slow modes (sometimes called ``vortex'' for $\omega_c(\bm{k})=0$) without variation along $z$; otherwise, they are fast modes (waves), in the terminology of the resonant interaction theory (Chen et al. 2005; Waleffe 1993; Greenspan 1990). \textit{In this sense, we have also found exact solutions on the slow manifold} which is well-known to reasonably decouple from the fast one at small Rossby number, at least in an intermediate transient state.

If we want to connect arbitrary-amplitude superposed modes to turbulence, randomness of $k_*$ and $c^*$ would be needed so that the ensemble could have modes of different wavelengths and chiralities. One possible way to understand such randomness is to imagine a field composed of such exact modes ``locally'' frozen in some randomly distributed space-time regions, a generalized sense of quenched disorder. Note that when the dissipation $e^{-\nu k_*^2t}$ is also included, such randomness of $k_*$ leads to strong intermittency phenomenology of turbulence dissipation due to the fact that small variations of $k_*$ will lead to strongly (Gaussian here) amplified fluctuations of dissipation, a mechanism proposed by Kraichnan (1967).

\subsubsection{The stratified vorticity mode}
The stratified vorticity mode
   \begin{equation}\label{eq:stratified}
        \bm{u}=\sum_{\bm{k}\times\bm{\kappa}=0} \hat{u}^c(\bm{k})\hat{\bm{h}}_c(\bm{k})e^{\hat{i}\bm{k}\cdot \bm{r}} e^{-\hat{i}\omega_c(\bm{k}) t}, %e^{-\nu k^2t}%,\ \bm{b}=\gamma\bm{u},
   \end{equation}
with fixed but arbitrary $\bm{\kappa}$, has been well discussed by Moses (1971), with even an example of ocean flows on a rotating earth and will not be further elaborated here.
But to be definite, here we also give a simple periodic example flow composed of three, or even more, modes with $\bm{\kappa}=(0,0,1)$:
\begin{eqnarray}\label{eq:XYz}
 \nonumber \dot{x} &=&u_x= X_1\cos(k_1z)+X_2\cos(k_2z)+X_3\cos(k_3z)+... \\
  \dot{y} &=&u_y= Y_1\cos(k_1z)+Y_2\cos(k_2z)+Y_3\cos(k_3z)+...,\\
 \nonumber \dot{z} &=&u_z= 0
\end{eqnarray}
with $X_i$, $Y_i$ and $k_i$ ($i=1,2,3,...$) being real numbers. [Of course, $u_x$ and $u_y$ in general can be arbitrary functions of $z$ and may have Kelvin-Helmholtz type of instability.] As said in the introductory discussion, we may call such a 1D2C mode the XYz flow.

We see from $\omega_c=c2\bm{k}\cdot\bm{\Omega}_0/k=c2\bm{\kappa}\cdot\bm{\Omega}_0$ that $\omega_c$ is constant now, and the wave speed $\omega_c/k$ depends on the chirality and wavelength. The necessary remark is that such a mode has $\bm{u}\perp \bm{\kappa}$. In each plan perpendicular to $\bm{\kappa}$, the velocity is uniform and only varies in the $\bm{\kappa}$ direction.
In other words, the flow has the interesting property that its velocity must vanish in one coordinate direction, say, $z$ as in Eq. (\ref{eq:XYz}), while the velocity components in the other two coordinate directions do not depend on their own coordinates but just the null-velocity coordinate.

Like the ABC flow or its generalization as explicitly given by us in the previous subsection, the 1D2C XYz mode, Eq. (\ref{eq:XYz}), or the damped and/or wave field with the time factor, is not Eulerian turbulence in the general sense, but we believe it can also show ``Lagrangian turbulence'' with sufficient number of Fourier modes and is unstable, whose detailed analytical and/or numerical studies, parallel to those of Dombre et al. (1986), Moffatt (1986) and Galloway and Frisch (1987), are however beyond the scope of this work.
However, since $\bm{\kappa}$ is fixed (though arbitrary), possible relevance to rotating turbulence might be made with the randomness of $\bm{\kappa}$ in all possible directions (though preferentially perpendicular to $\bm{\Omega}_0$ as found under the framework of resonant wave theory). This randomness, again, may come from the exact solutions locally frozen in some (random) space-time regions.

\subsection{Rotating Magnetohydrodynamics}
Chandrasekhar (1951; Secs. 40, 49) already treated MHD without and with rotation, the former with two exact solutions (Alfvenian and force-free ones).

For incompressible single-fluid MHD in the rotating frame
the dynamics read
\begin{eqnarray}\label{eq:mhd}
  \partial_t\nabla\times\bm{u} &=& \nabla\times[\bm{u}\times(\nabla\times\bm{u})]-\nabla\times[\bm{B}\times(\nabla\times\bm{B})]+2\bm{\Omega}_0\cdot\nabla\bm{u} \label{eq:mhd1}\\
  \partial_t\bm{B} &=& \nabla\times(\bm{u}\times\bm{B}).\label{eq:mhd2} %\nonumber
\end{eqnarray}
When in the inertial frame ($\bm{\Omega}_0=0$), the linear dynamics (and that the dispersion relation $|\omega|=|\bm{k}\cdot\bm{B}_0|$) of the equation, after introducing $\bm{B}=\bm{B}_0+\bm{b}$, requires that $\bm{u}=\pm\bm{b}$ which happens to kill the nonlinearity without any other requirements on the superposition of helical modes. Since the helical modes form the complete bases for any function (in $L^2$ space, say), the ansatz of the wave can be arbitrary. This is the well-known nonlinear Alfv\'en wave. When in the rotating frame with the Coriolis term $2\bm{\Omega}_0\cdot\nabla\bm{u}$ working in the first equation, the linear dispersion relation no longer guarantees the cancelation of the nonlinear terms and that only special superposition of the helical modes, the mono-wavelength homochiral vorticity mode and the stratified vorticity mode, are exact solutions.

\subsubsection{Mono-wavelength homochiral Beltrami mode}
We apply $\bm{U}={\bm u}, %\bm{E}=\bm \varepsilon,
{\bm B}={\bm B}_0+\bm b%, P_s=P_{s0}+p_s
$ to derive the dispersion relation.
Rewriting MHD equations with these variables, just as our rotating fluids case due to $\bm{\Omega}_0$, it is familiar that an extra linear term emerges (Moffatt 1978, Sec. 10.2). Things are similar to all that in the rotating fluids except that we now have an extra variable $\bm{b}$ which is easy to be taken care of: The mono-wavelength homochiral solution is
\begin{eqnarray}
\nabla\times\bm{u} &=& k_*\bm{u} \label{eq:ualignment}%\\
%\bm{b} &=& \gamma\bm{u} \label{eq:ubalignment}
\end{eqnarray}
with $\bm{u}$ given by Eq. (\ref{eq:mhU}) (\textit{force-free}, both for Lorentz and Magnus).
Applying the helical wave representation, we can derive the dispersion relation more straightforwardly as follows. We replace $\bm{u}$ in the corresponding linear equations
\begin{eqnarray}
  \partial_t\nabla\times\bm{u} &=& -\nabla\times[\bm{B}_0\times(\nabla\times\bm{b})]+2(\bm{\Omega}_0\cdot\nabla)\bm{u} \label{eq:lmhd1}\\
  \ \partial_t\bm{b} &=& \nabla\times(\bm{u}\times\bm{B}_0) \label{eq:lmhd2}%\nonumber
\end{eqnarray}
with $\hat{\breve{\bm{u}}}^c(\bm{r}|\bm{k})$ defined by
\begin{equation}\label{eq:dspu}
\hat{\breve{\bm{u}}}^c(\bm{r}|\bm{k})\triangleq \hat{u}^c(\bm{k})\hat{\bm{h}}_c(\bm{k})e^{\hat{i}\bm{k}\cdot\bm{r}}e^{-\hat{i}\omega_c(\bm{k}) t},
\end{equation}
and similarly for $\bm{b}$, which leads to:
\begin{eqnarray}\label{eq:lmhdw}
  (2\bm{\Omega}_0\cdot\bm{k}+c\omega_c k)\hat{u}^c+ck\bm{B}_0\cdot\bm{k}\hat{b}^c &=& 0\label{eq:lmhdw1}\\
  \bm{B}_0\cdot\bm{k}\hat{u}^c+\omega_c\hat{b}^c &=& 0,\label{eq:lmhdw2}
\end{eqnarray}
which %, combined with Eq. (\ref{eq:ubalignment}),
give
\begin{equation}\label{eq:gamma}
\tilde{\gamma}(\bm{k})=\frac{\hat{b}_c}{\hat{u}_c}=-\frac{\bm{B}_0\cdot\bm{k}}{\omega_c}=-\frac{2\bm{\Omega}_0\cdot\bm{k}+c\omega_c k}{ck\bm{B}_0\cdot\bm{k}}.
\end{equation}
Here we have explicitly designated the frequency with the chirality index ``$c$'', determined by $\omega_c(2\bm{\Omega}_0\cdot\bm{k}+c\omega_c k)-ck(\bm{B}_0\cdot\bm{k})^2=0$ from Eqs. (\ref{eq:lmhdw1}) and (\ref{eq:lmhdw2}): $$\omega_c=[-\bm{\Omega}_0\cdot\bm{k}\pm\sqrt{(\bm{\Omega}_0\cdot\bm{k})^2+k^2(\bm{B}_0\cdot\bm{k})^2}]/(ck).$$

There are some conditions that should be satisfied to make such mono-wavelength homochiral waves be exact arbitrary-amplitude solutions. First of all, $\bm{B}_0\cdot\bm{k}$ should not be zero. Also, in the above, we have assumed that the nonlinearity in the induction equation (\ref{eq:mhd2}) vanishes, which however requires extra condition(s), say, $\bm{\Omega}_0\cdot\bm{k}/\Omega_0=\bm{B}_0\cdot\bm{k}/B_0$ which leads to
\begin{equation}\label{eq:ubalignment}
\bm{b}=\gamma \bm{u}
\end{equation}
with constant $\gamma=\tilde{\gamma}(\bm{k})$: now $\bm{k}$ is further restricted in the symmetric plane spanning equal angles with $\bm{\Omega}_0$ and $\bm{B}_0$. Another special restriction on the superposition will be discussed a bit later.

\subsubsection{Extension of Moses' stratified vorticity mode}
For stratified vorticity mode of rotating MHD, the only necessary remark is that both $\bm{b}$ and $\bm{u}$ share the same $\bm{\kappa}$, as can be seen from Eqs. (\ref{eq:lmhdw1}) and (\ref{eq:lmhdw2}): Like the mono-wavelength homochiral Beltrami mode in the last subsection, the relation between $\bm{b}$ and $\bm{u}$ is determined by the linear dynamics, i.e., the same dispersion relation. But now, to kill the nonlinearity in Eq. (\ref{eq:mhd2}), we do not need extra conditions, as explained in the introductory discussion (Sec. 1.3, with $\nabla \times \bm{u}$ replaced by $\bm{b}$ here). So, the extension to rotating MHD of Moses' stratified mode is more robust than the extension of Beltrami mode.

Unlike the Beltrami mode, such a mode, Eq. (\ref{eq:stratified}), does not have clear parity-violation signature (though could be imposed by restricting to homochiral modes, say); thus no clear conventional connection to alpha dynamo. From Eqs. (\ref{eq:lmhdw1}) and (\ref{eq:lmhdw2}), in the direction $\bm{\kappa}\perp\bm{B}_0$ and $\bm{\kappa}\perp\bm{\Omega}_0$, the waves are, respectively, the same as the inertial wave of neutral rotating fluids and Alfv\'en wave of MHD in the inertial frame. Note that for Alfv\'en wave $\omega$ does not depend on the chirality ``$c$'' of the wave, so we can not associate downward or upward propagation to the waves of positive or negative helicity, unlike the inertial wave as discussed by Moffatt (1978).
Just as Moses (1971) did for the ocean flows on a rotating earth, it would be interesting to find explicit examples, with further specifications of Eq. (\ref{eq:stratified}) or more explicitly of the 1D2C formula (\ref{eq:XYz}) of such a mode relevant to astrophysical rotating MHD objects, which is however beyond the scope of this note.

\subsubsection{Special superposition principles and the role of nonlinearity in turbulence}
As long as $\bm{\Omega}_0\times\bm{B}_0\ne 0$, typically in the Earth's core where the differential rotation between the inner core and the mantle is believed to be strong enough to cause the magnetic field to be mainly toroidal, locally perpendicular to the rotation axis, we can find a $\bm{k}$-plane on which $\bm{k}\cdot\bm{\Omega}_0=0$, and where we see from Eqs. (\ref{eq:lmhdw1}) and (\ref{eq:lmhdw2}) that the dispersion relationship leads again to $\bm{u}=\pm\bm{b}$, reducing to the classical nonlinear Alfv\'en wave. \textit{That is, for the rotating MHD, there is yet a third superposition principle of nonlinear waves: In the $\bm{k}$ plane perpendicular to $\bm{\Omega}_0$, arbitrary amplitude waves of arbitrary wavelengths and either chirality can be superposed.} The only difference to the classical pure Alfv\'en wave is that now $\bm{k}$ is restricted to this plane and that the harmonic modes are not complete (for the $L^2$ space), thus no completely ``arbitrary'' shapes of the waves. And, of course, this plane contains only slow modes when $\bm{\Omega}_0\times\bm{B}_0= 0$. This specific example highlights the fact that, though we have proved the only two generic exact solutions killing the quadratic nonlinearity common to various 3D incompressible hydrodynamic type models, there are other interesting ``accidental'' superposition principles for specific situations.

Now we would like to add some remarks on the rotating MHD turbulence where the exact nonlinear wave solutions might play some role. \citet{FavierGodeferdCambonGAFD11} performed both linear and nonlinear numerical studies, but only for $\bm{\Omega}_0\times\bm{B}_0= 0$. Of course, the data show that the alignment between $\bm{u}$ and $\bm{b}$ is randomly distributed, even for $\bm{\Omega}_0=0$ (where the pure Alfv\'en waves present exact alignment and equipartition), though indeed the alignment is stronger for weaker rotation. Such a consistency may be better connected to our nonlinear exact wave solutions rather than to the linear arguments by those authors; nevertheless, the relevance of the nonlinearity-killing exact solutions with turbulence is not very clear yet. Actually, it is also illuminating to relate the turbulence data to the opposite extreme, the statistical absolute equilibrium purely coming from the nonlinear interactions. For instance, regarding the ``slight excess of magnetic energy with respect to the kinetic one for inertial and dissipative scales'' in \citet{FavierGodeferdCambonGAFD11}'s Fig. 9a, we may connect it to the same excess in the absolute equilibrium spectra: We easily see from Eq. (2.15) of Zhu, Yang \& Zhu (2014) that magnetic energy exceeds kinetic energy for every $k$, as the sum over two chiral sectors (but not for each sector). At least part of the solutions of turbulence must maintain the nonlinearity to have a cascade/transfer, as was also pointed out by Moses (1971), while exact nonlinearity-kill solutions, like the pure nonlinear Alfv\'en waves, are also allowed; thus a mixture.

\subsection{On Hall and fully two-fluid MHD}
We are now ready to address the Hall and two-fluid MHD (where helical waves already exist without the necessity of other mechanisms such as rotation) on whose exact solutions some disagreements have arosen among several groups (Mahajan \& Krishan 2005, Galtier 2006, Sahraoui et al. 2007, Mahajan \& Miura 2009, Verheest 2009). Interested readers should refer to those articles for more details, as the purpose of this subsection is only to help clarify the controversies using our systematic treatment as established in the previous (sub)sections.

The dynamics state that the generalized vorticities are ``frozen in'' to the flows and, in Alfv\'en unit, read
\begin{equation}\label{eq:HMHD}
\partial_t \bm{\Omega}_s=\nabla\times(\bm{U}_s\times\bm{\Omega}_s).%\cancel{+\lambda\Delta\bm{\Omega}_s}
\end{equation}
For full two-fluid MHD, we have the canonical momenta $\bm{P}_s=m_s\bm{U}_s+q_s\bm{A}$ ($m_s$ and $q_s$ being the mass and charge respectively) and that the generalized vorticities $\bm{\Omega}_s=m_s\nabla\times\bm{U}_s+q_s\bm{B}$ (frozen in to $\bm{U}_s$). Since now $\bm{u}_s$ depends on the species, naturally we can have species-dependent $k_{*s}$ and $\gamma_s$, in the fashion of Eqs. (\ref{eq:mhU}), (\ref{eq:ualignment}) and (\ref{eq:ubalignment}).
For the dispersion relations, $\omega_c$, can also be derived as we did for rotating MHD (with the help of the appropriate Maxwell equation).
We use $\bm{U}_s={\bm u}_s, %\bm{E}=\bm \varepsilon,
{\bm B}={\bm B}_0+\bm b%, P_s=P_{s0}+p_s
$ with $\nabla\times\bm{B}_0=0$. The two-fluid ``frozen-in'' equations then become
\begin{equation}\label{eq:decomposed}
\partial_t (m_s\nabla\times\bm{u}_s+q_s\bm{b})=\nabla\times[\bm{u}_s\times(m_s\nabla\times\bm{u}_s+q_s\bm{b})]+q_s \nabla\times(\bm{u}_s\times \bm{B}_0)
\end{equation}
which, as said, is solved by the Beltrami wave (to which we are limiting ourselves for the time being), Eq. (\ref{eq:mhU}), i.e.,
$$\bm{u}_s=[ \sum_{|\bm{k}|=k_s,c=c^*_s} \hat{u}_s^c(\bm{k})\hat{\bm{h}}_c(\bm{k})e^{\hat{i}\bm{k}\cdot \bm{r}} ] e^{-\hat{i}\omega t},%\ \bm{b}=\alpha_s\bm{u}_s
$$
with $c^*_s$ being uniformly $+$ or $-$ and $k_s$ constants%; similarly, the extension of the stratified vorticity mode of Moses, Eq. (\ref{eq:stratified})
. The dispersion relation can be obtained in the conventional way (Stix 1992)
or more straightforwardly by using helical representation from the beginning as follows, which will also settle down the relations among $k_s$ and $\alpha_s$ with $\bm{b}=\alpha_s\bm{u}_s$. The dispersion relation is determined by the Maxwell equations and the linear part of Eq. (\ref{eq:decomposed}), that is
\begin{equation}\label{eq:linear}
    \partial_t (m_s\nabla\times\bm{u}_s+q_s\bm{b})=q_s \nabla\times(\bm{u}_s\times \bm{B}_0).
\end{equation}
From the above helical representation, as in the conventional derivation of dispersion relation with mono-wavelength, but now also uni-chiral wave, we replace in the above equation $\bm{u}_s$ with Eq. (\ref{eq:dspu}), i.e.,
$$\hat{\breve{\bm{u}}}_s^c=\hat{u}_s^c(\bm{k})\hat{\bm{h}}^c(\bm{k})e^{\hat{i}(\bm{k}\cdot\bm{x}-\omega t)},$$
and similarly for $\bm{b}$, which from the above linear equation (\ref{eq:linear}) and $\nabla\times\bm{b}=\mu_0\sum_s q_s n_s\bm{u}_s \cancel{ + \partial_t \bm{E}}$ (for simplicity, we have neglected the displacement current $\partial_t \bm{E}$ which of course could be included in the calculation for more general results), leads to:
$$\omega[cm_sk\hat{u}_s^c(\bm{k})+q_s\hat{b}^c(\bm{k})]=-q_sB_0k_{\parallel} \hat{u}_s^c(\bm{k}), \ ck\hat{b}^c(\bm{k})=\mu_0\sum_s q_sn_s\hat{u}_s^c(\bm{k}).$$
Here $k_{\parallel}=\bm{k}\cdot \bm{B}_0/B_0$. Solving $\omega$ we find $\bm{b}=\alpha_s\bm{u}_s$ with $\alpha_s$ indeed a constant, depending only on $k$ (but not $k_{\parallel}$) which is fixed for the Beltrami mode. Since the dynamics of all species share the same $\bm{b}$, there should be only a unique, though arbitrary, wavenumber $k$ on which shell of wavevectors can we have the mono-wavelength homochiral superposition to form such an exact solution: It appears that Verheest's (2009) remark, pointing to the claim of Mahajan \& Krishan (2005), that one could have as many number of $k$s as there are species making up the plasma, is still incomplete in the sense that the requirement of $k$ being real and positive was not considered there.
The ideal incompressible Hall MHD has $\bm{\Omega}_s=\nabla\times\bm{P}_s$, with canonical momenta $\bm{P}_i=\bm{A}+\bm{U}$ and $\bm{P}_e=\bm{A}$ for each species $s$, ``frozen in'' to the respective flows $\bm{U}_i=\bm{U}$ and $\bm{U}_e=\bm{U}-\nabla\times\bm{B}$ with $\bm{B}=\nabla\times\bm{A}$. Again, taking $\bm{U}=\bm{U}_0+\bm{u}$ and $\bm{B}=\bm{B}_0+\bm{b}$, we have the dispersion relation for Hall MHD as given by Mahajan and Krishan (2005), and by equating the relationship determined by the dispersion relation with $\bm{b}=\gamma\bm{u}$ [with which either of the linear parts of their equation 6 or 7 gives $\omega=(ck-\gamma^{-1})k_{\parallel}$], it is seen that our $\gamma$ is equal to their $\alpha$. Such a procedure provides a solution more general than their ``most general three-dimensional solution'' [c.f., Eqs. (\ref{eq:ABCD}) and (\ref{eq:ABCDEF}) in our Sec. \ref{app:ABC}.] Unlike the classical single-fluid MHD, there is no necessity of other conditions to make the extended Beltrami wave a solution.

The extension of the stratified vorticity mode of Moses, Eq. (\ref{eq:stratified}) is also straightforward, just as that in the classical one-fluid MHD (with the same $\bm{\kappa}$ shared by all).
In the context of two-fluid MHD, Mahajan and Miura (2009) and Verheest (2009) also noticed the stratified vorticity mode: Their derivations, without the formulation of helical representation, unfortunately appear to have led to some confusion and disagreements in understanding the results among them, and now we hope our unified treatment can help make it clearer, especially with the explicit example of such 1D2C mode given in Eq. (\ref{eq:XYz}).

Note that in incompressible dynamics, the velocity (momentum) and magnetic (Faraday's law) dynamics are decoupled from the energy or state equation. Galtier (2006) and Sahraoui et al. (2007) argued that Hall MHD transverse-field dynamics was not compatible with the state equation, specifically the pressure closed by the polytropic state equation, and that the statement of Hall MHD exact solution by Mahajan \& Krishan (2005) was ``not correct'' (Galtier 2006); they claimed that incompressible Hall MHD only makes sense in the large-beta (thermal/magnetic pressure ratio) limit. Such physical discussions on the consistency of the incompressibility approximation with extra (state) models are interesting but have nothing to do with our methodology and are beyond the scope of this work.

Again, these exact nonlinear wave solutions with ``quenched'' (in space and/or time) disorder in the parameters ($k_*$, $\gamma$ and $\alpha_s$) might help understand turbulence. In other words, though the $k_*$ and $\gamma$ are fixed for each Beltrami mode as also correctly pointed out by Verheest (2009), it may still be possible to salvage the attempt to relate the scale-dependent alignment between $\bm{b}$ and $\bm{u}$ to the solar atmosphere turbulence (Krishan \& Mahajan 2004, Mahajan \& Krishan 2005). Indeed, in a turbulent state, instead of global alignment with the fixed parameter as given by the exact solutions shown here, fluctuations can be related to local alignments for each $\bm{k}$, in the nonlinear vortex sense (Zhu et al. to appear) beyond that of linear waves (Meyrand \& Galtier 2012), and one-chiral-sector-dominated states (Zhu, Yang \& Zhu  2014), both of which are in the statistical sense but not for any single exact solution.

\section{Discussions}
An important purpose and, indeed, the consequence of this cross-disciplinary study is to obtain universal insights about a series of hydrodynamic type equations governing the dynamics of various materials in nature (atmospheres, oceans, solar winds, accretion disks etc.), in industry and laboratory (turbomachinery, fusion facilities such as the International Thermonuclear Experimental Reactor, ......)
We have systematically discussed the exact nonlinear wave solutions of a series of hydrodynamic type models as the superpositions of helical waves. The Beltrami mode and the stratified vorticity mode are proved to be the only generic superposition principles, though other ``accidental'' superposition principles due to special combinations of the nonlinear terms are possible. Relevant knowledge was already partially noticed at different situations by different authors (see, e.g., Beltrami 1889, Moses 1971, Moffatt 1978, Dombre et al. 1986, Verheest \& Das 1989, Mahajan \& Krishan 2005, Mahajan \& Miura 2009 and Verheest 2009) but are lack of completeness and/or unambiguity, which has led to some confusion and controversies, and, which now are settled. The novelty of the approach also lies in the helical superposition formulation and in the generalization and extension of exact vortex solutions to waves.

The dispersion relation of the waves for the classical single-fluid MHD with a uniform background in the inertial frame leads to $\bm{u}=\bm{b}$ which naturally kills the nonlinearities, thus a nonlinear solution of arbitrary amplitude. The rotation effect (Coriolis force) imposes nontrivial constraints on the superposition principle: The dispersion relation in general keeps the nonlinearities and leads to further restrictions (to special manifolds of wavevector space) on the Beltrami waves and Alfv\'en waves. Such strengthened restriction is quite special compared to rotating neutral fluids, Hall and two-fluid MHDs, giving us a lesson about the specificity.
Our result may be useful for dynamo study [see also the last remark in (i) of Sec. \ref{app:ABC} concerning ABC$def...$ flows]; also, it's promising to fit the ideas to more realistic systems such as the important rotating cylindrical Hall MHD (Krishan \& Varghese 2008).

What exactly are the physical roles of such exact solutions in turbulence issues (such as intermittency) is an intriguing question. Except that the exact solutions, such as the Alfv\'en waves, are frequently used for reference (see, e.g., Boldyrev 2006), there has not been much clear evidence. But nevertheless, instability (Moffatt 1986; Galloway \& Frisch 1987) and absolute equilibrium (Zhu, Yang \& Zhu 2014; Zhu 2014) considerations with the stable Gibbs ensemble (Kraichnan 1959) do offer some clues; and, we have introduced (quenched) disorder in the parameters of the exact solutions, which might help to bridge such solutions and the complexity of turbulence. On the other hand, the absence of the nonlinear interactions of the exact-solution modes leads to zero contribution to the transfers of whatever the restricted ``chiroid'' ensembles, such as those specific absolute equilibrium Gibbs ensembles of Zhu, Yang \& Zhu (2014) and Zhu (2014). Of course, restricting to only the exact-solution modes would lead to no thermalisation mechanism for the absolute equilibrium, i.e., depletion of nonlinearity, so in this sense, this work complements those previous studies in understanding the nonlinearities in another way.

\section*{Acknowledgments}
The author acknowledges the interactions with Prof. M. Y. Yu and Messrs. Z.-W. Xia and H.-S. Xie.

% susie put cite commands here, don't bother with citet etc just yet.

\bibliographystyle{jfm}

\bibliography{exactJFM2}

\begin{thebibliography}{14}
\expandafter\ifx\csname natexlab\endcsname\relax\def\natexlab#1{#1}\fi

\bibitem[\protect\citeauthoryear{Alfv\'en \& F\''althammer 1963}{1963}]{af63}
{\sc Alfv\'en H. \& F\"althammer C.} 1963 Cosmical Electrodynamics. Clarendon
Press, Oxford

\bibitem[\protect\citeauthoryear{Beltrami 1889}{1889}]{Beltrami1889}
{\sc Beltrami E.} 1985 Condideratioins on hydrodynamics Int. J. Fusion Energy, 3, 53--57: Translated from the original 1889 paper into English by Filipponi G.

\bibitem[Biferale, Musacchio \& Toschi(2012)]{bmt12}
{\sc Biferale, L., Musacchio, S. \& Toschi, F.} 2012 Inverse energy cascade in three-dimensional
isotropic turbulence. Phys. Rev. Lett. {\bf 108}, 104501¨C104504.

%\bibitem[\protect\citeauthoryear{Biferale et al. 2013}{2013}]{bmt13}
%{\sc Biferale et al.} 2013 Split energy-helicity cascades in
%three-dimensional homogeneous and isotropic
%turbulence. J. Fluid Mech., {\bf 730}, 309--327.

\bibitem[\protect\citeauthoryear{Boldyrev 2006}{2006}]{b06}
{\sc Boldyrev S.} 2006, Astrophysical Journal, 626, L37.

%\bibitem[\protect\citeauthoryear{Bourouiba 2008}{2008}]{b08}
%{\sc Bourouiba L.} 2008 Model of a truncated fast rotating flow at infinite Reynolds number. Phys. Fluids, 20, 075112.

\bibitem[\protect\citeauthoryear{Cambon \& Jacquin 1989}{1989}]{cj89}
{\sc Cambon C., Jacquin L.} 1989 Spectral approach to non-isotropic turbulence
subjected to rotation. J. Fluid Mech., 202, 295--317.

\bibitem[\protect\citeauthoryear{Chandresekhar 1951}{1951}]{c51}
{\sc Chandrasekhar S.} 1951 Hydrodynamic and hydromagnetic stability. Oxford University Press

\bibitem[Chen et al.(2005)]{CCEH05}
{\sc Chen, Q. N., Chen, S. Y., Eyink, G. L. \& Holm, D.} 2005 Resonant
interactions in rotating homogeneous three-dimensional
turbulence. {\em J. Fluid Mech.\/} {\bf 542} (2), 139--163.

\bibitem[Davidson(2013)]{DavidsonBook13}
{\sc Davidson, P.} 2013 Turbulence in Rotating, Stratified and Electrically Conducting Fluids. Cambridge University Press.

%\bibitem[Dhar \& Lebowitz(2008)]{DharLebowitz}
%Dhar, D., Lebowitz, J. L. 2008 Restricted equilibrium ensembles: Exact equation of state of a model glass.
%{\em Europhysics Letters} {\bf 92}, 20008.

\bibitem[\protect\citeauthoryear{Dombre et al.}{1986}]{DombreJFM86}
{\sc Dombre T. et al.} 1986 Chaotic streamlines in the ABC flows. J. Fluid Mech. {\bf 167} 353--391.

\bibitem[Favier, Godeferd \& Cambon(2011)]{FavierGodeferdCambonGAFD11}
{\sc Favier, B. F. N., Godeferd, F. S. \& Cambon, C.} 2011 On the effect of rotation on magnetohydrodynamic turbulence at
high magnetic Reynolds number. Geophysical and Astrophysical Fluid Dyanmics. 1--23.

\bibitem[Galloway \& Frisch (1987)]{GallowayFrisch87}
{\sc Galloway D. \& Frisch U.} 1987 A note on the stability of a family of space-periodic Beltrami flows. J. Fluid Mech. {\bf 180} 557--564.

\bibitem[\protect\citeauthoryear{Galloway 2012}{2012}]{g12}
{\sc Galloway D.} 2012 ABC flows then and now. Geophysical and Astrophysical Fluid Dynamics, {\bf 106}, 450.

\bibitem[\protect\citeauthoryear{Galtier 2006}{2006}]{g03}
{\sc Galtier S.} 2003 Weak inertial-wave turbulence theory. Phys. Rev. E, {\bf 68}, 015301(R)

\bibitem[\protect\citeauthoryear{Galtier 2006}{2006}]{g06}
{\sc Galtier S.} 2006 Wave turbulence in incompressible Hall
magnetohydrodynamics. J. Plasma Phys., 72, 721--769.

\bibitem[\protect\citeauthoryear{Greenspan 1990}{1990}]{g68}
{\sc Greenspan H. P.} 1990 The theory of rotating fluids. Breukelen Press

\bibitem[\protect\citeauthoryear{Krishan \& Mahajan 2004}{2004}]{k04}
{\sc Krishan V. \& Mahajan S.} 2004 Hall-MHD turbulence in the solar atmosphere. Sol. Phys., 220, 29--41.

\bibitem[\protect\citeauthoryear{Krishan \& Varghese 2008}{2008}]{kv08}
{\sc Krishan V., Varghese B. A.} 2008 Cylindrical Hall-MHD Waves: A Nonlinear Solution. Sol. Phys., 247, 343--349.

\bibitem[Kraichnan(1959)]{k59}
{\sc Kraichnan R. H.} 1959 Classical Fluctuation-Relaxation Theorem. Phys. Rev., {\bf 113}, 1181--1182.
%
%\bibitem[\protect\citeauthoryear{Kraichnan}{1967a}]{k67a}
%{\sc Kraichnan R. H.} 1967a Inertial ranges in two-dimensional turbulence. Phys. of Fluid, 102, 1417

\bibitem[\protect\citeauthoryear{Kraichnan}{1967}]{k67}
{\sc Kraichnan R. H.} 1967 Intermittency in the Very Small Scales of Turbulence. Phys. of Fluid, {\bf 10}, 2080--2081.

\bibitem[\protect\citeauthoryear{Mahajan \& Krishan 2005}{2005}]{mk05}
{\sc Mahajan S. M., Krishan V.} 2005 Exact solution of the incompressible Hall magnetohydrodynamics. MNRAS, 359, L27--L29.

\bibitem[\protect\citeauthoryear{Mahajan \& Miura}{2009}]{mm09}
{\sc Mahajan S. M., Miura H.} 2009 Linear superposition of nonlinear waves. J. Plasma Phys., 75, 145--152.

\bibitem[\protect\citeauthoryear{Meyrand \& Galtier 2012}{2012}]{mg12}
{\sc Meyrand R. \& Galtier S.} 2012 Spontaneous Chiral Symmetry Breaking of Hall Magnetohydrodynamic Turbulence. Phys. Rev. Lett., 109, 194501.

%\bibitem[\protect\citeauthoryear{Mininni et al. 2005}{2005}]{m05}
%{\sc Mininni P. et al.} 2005 Direct Simulations of Helical Hall-MHD Turbulence and Dynamo Action. Astrophysical J., 619, 1019

\bibitem[\protect\citeauthoryear{Moffatt}{1978}]{Moffatt}
{\sc Moffatt H. K.} 1978 Magnetic Field Generation in Electrically Conducting Fluids. Cambridge University Press

\bibitem[Moffatt(1986)]{Moffatt}
{\sc Moffatt H. K.} 1986 Magnetostatic equilibria and analogous Euler flows of arbitrarily complex topology. Part 2. Stability considerations. 
J. Fluid. Mech., {\bf 166}, 359--378.

\bibitem[\protect\citeauthoryear{Moses 1971}{1971}]{m71}
{\sc Moses H. E.} 1971 Eigenfunctions of the Curl Operator, Rotationally Invariant Helmholtz Theorem, and Applications to Electromagnetic Theory and Fluid Mechanics. 
SIAM J. Appl. Math., 21, 114--144.

\bibitem[Sagaut \& Cambon(2008)]{SagautCambon08book}
{\sc Sagaut, P. \& Cambon C.} 2008 Homogeneous Turbulence Dynamics. Cambridge University Press.

\bibitem[\protect\citeauthoryear{Sahraoui et al. 2007}{2007}]{s07}
{\sc Sahraoui et al.} 2007 On waves in incompressible Hall
magnetohydrodynamics. J. Plasma Phys., 73, 723--730.

\bibitem[Stix (2007)]{StixBook92}
{\sc Stix, T. H.} 1992 Waves in Plasmas. American Institute of Physics.

\bibitem[\protect\citeauthoryear{Verheest 1999}{1999}]{v09}
{\sc Verheest F.} 2009 Linear description of nonlinear electromagnetic cold plasma modes based on
generalized vorticity. Phys. Plasmas, 16, 082104.

\bibitem[\protect\citeauthoryear{Verheest \& Das}{1989}]{vd89}
{\sc Verheest F., Das K. P.} 1989 Finite-amplitude circularly polarized waves in a magnetized multispecies plasma with drifts. 
Plasma Phys. and Controled Fusion, {\bf 31}, 103.

\bibitem[Waleffe(1992)]{W92}
{\sc Waleffe, F.} 1992 The nature of triad interactions in homogeneous turbulence.
{\em Phys. Fluids A\/} {\bf 4}, 350--363.

\bibitem[Waleffe(1993)]{W93}
{\sc Waleffe, F.} 1993 Inertial transfers in the helical decomposition.
{\em Phys. Fluids A\/} {\bf 5}, 677--685.

\bibitem[\protect\citeauthoryear{Zhu et al.}{2014}]{z14}
{\sc Zhu, J.-Z., Yang, W. \& Zhu, G.-Y.} 2014 Purely helical absolute equilibria and chirality of
(magneto)fluid turbulence. J. Fluid Mech., {\bf 739}, 479--501.

\bibitem[Zhu(2014)]{Zhu14}
{\sc Zhu, J.-Z.} 2014 Note on specific chiral ensembles of statistical hydrodynamics: ``Order function'' for
transition of turbulence transfer scenarios. Phys. Fluids, {\bf 26}, 055109.

%\bibitem[Zhu et al. (2014)]{ZhuETC14}
%{\sc Zhu, J.-Z. et al.} 2014 Chirality of sub-ion plasma turbulence. To appear.


\end{thebibliography}

\end{document}